\documentclass[twocolumn,prb,groupedaddress,showpacs,superscriptaddress]{revtex4-2}

\usepackage{amsmath,amssymb,amsfonts}
\usepackage{algorithmic}
\usepackage{graphicx}
\usepackage[caption=false,font=footnotesize]{subfig} 
\usepackage{textcomp}
\usepackage{xcolor}
\usepackage{listings}
\usepackage{hyperref}
\usepackage{url}

\newcommand{\CM}{{\mathbb C}}

\newcommand{\Hh}{{\mathcal H}}

\begin{document}

\title{Topological Obstructions in Quantum Adiabatic Algorithms}

\author{Prathamesh S. Joshi}
\email{pjoshi1@mail.yu.edu}
\address{Department of Computer Science, Yeshiva University, New York, NY 10016, USA
}

\author{Emil Prodan}
\email{prodan@yu.edu}
\address{Department of Physics, Yeshiva University, New York, NY 10016, USA
}

\begin{abstract} We point out that, when an optimization problem has more than one solution, the quantum adiabatic algorithms (QAA) encounter topological obstructions leading to adiabatic spectral flows where spectral branches unavoidably traverse the spectral gap above the ground states of the quantum Hamiltonians. This raises serious doubts about the validity of the algorithms in such situations. However, using the Max-Cut problem as an example, we explain and demonstrate here that QAAs correctly detect all existing solutions in one single run. This newly discovered capacity of QAAs to simultaneously  detect multiple solutions to an optimization problem can have an important impact on future developments of quantum variational algorithms.
\end{abstract}

\maketitle

\section{Introduction}

\begin{figure}[b!]
    \centering   
    \includegraphics[width=0.99\linewidth]{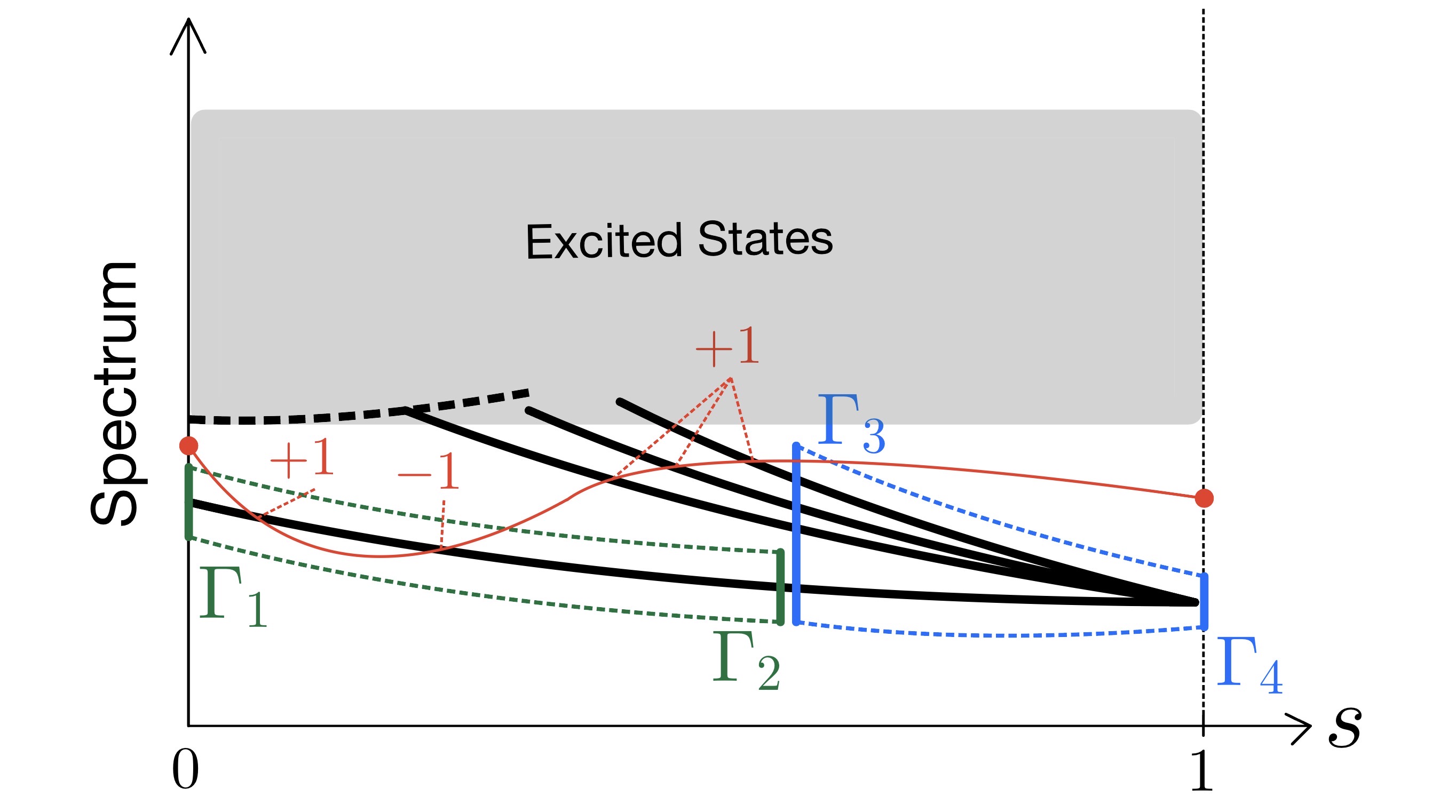}
    \caption{Schematic of the adiabatic spectral flow for the interpolation between an elementary Hamiltonian $H(0)$ and the Max-Cut Hamiltonian $H(1)$, in the case of multiple solutions. Overlaid are two pairs of spectral intervals $(\Gamma_1, \Gamma_2)$ and $(\Gamma_3,\Gamma_4)$, which can be deformed into each other as required by the Adiabatic Theorem (see the dotted lines). }
    \label{Schematic}
\end{figure}

The Adiabatic Theorem is a rigorous mathematical statement \cite{KatoJPSJ1950,AvronCMM1987} about a smooth family of Hamiltonians $\{H(s)\}$, $s\in [0,1]$, which can be  thought of as a deformation of an initial Hamiltonian $H(0)$ into a final one $H(1)$. Such a deformation can happen with various speeds in real time $t$, which is reflected by the time-dependent Hamiltonian $H(t/\tau)$. The Adiabatic Theorem applies in the regime of slow deformations, thus, for $\tau \gg 1$. When we enumerate the eigenvalues of $H(s)$ as $\lambda_n(s)$, we include the degeneracies and we assume that a complete orthonormal system of eigenvectors $\psi_n(s)$ has been chosen. Then, if $P_\Gamma(s)$ is the spectral projection of $H(s)$ onto the interval $\Gamma$ of the real axis,
\begin{equation}
    P_\Gamma(s) = \sum_{\lambda_n(s) \in \Gamma} |\psi_n(s)\rangle \langle \psi_n(s)|,
\end{equation}
the unitary time-evolution operator $U_\tau(t,t')$ generated by $H(t/\tau)$ has the property that, up to errors that go to zero as $\tau \to \infty$,
\begin{equation}
  U_\tau(t,0)  P_\Gamma(0) U_\tau(t,0)^\dagger = P_{\Gamma'}(t/\tau),
\end{equation}
provided $\Gamma$ can be deformed into $\Gamma'$ such that the ends of the interval avoid any spectrum of $H(s)$ (see Fig.~\ref{Schematic}). This is the statement of the Adiabatic Theorem \cite{KatoJPSJ1950,AvronCMM1987}.

If the conditions of the Adiabatic Theorem are met and the spectrum inside $\Gamma$ consists of a non-degenerate ground level, then the adiabatic time evolution gives us the means to reach the unique ground state of $H(1)$. Quantum Adiabatic Algorithms (QAA) take advantage of this mechanism by encoding the solution of an optimization problem into the ground state of $H(1)$. Specifically, the various options tested by the optimization process are encoded into a basis of the Hilbert space and $H(1)$ is designed such that its application on this basis returns the cost estimator which is to be minimized. Exact implementations of QAAs are costly in practice, and QAAs are rather used as guiding-function generators for more efficient variational quantum algorithms (VQA) \cite{BlekosPR2024,GrangeAOR2024}.

In this paper, we focus exclusively on fundamental aspects of QAAs. As such, we will use quantum simulators to demonstrate and assess their accuracy and performance, leaving the issues related to efficiency and implementation on quantum hardware out of our discussions, though we will analyze runs under simulated noise.  Among the many settings where QAAs were put to work, the Max-Cut problem definitely stands out. Indeed, it was among the first concrete real-world problems to be formulated as a VQA \cite{FarhiArXiv2014} and, these days, Max-Cut problem supplies a standard benchmark for VQA development. We recall that, for    a graph with set of vertices $V$, Max-Cut problem asks to partition $V = V_1 \cup V_2$ such that there is a maximum number of edges traversing between $V_1$ and $V_2$. This task is known to be an NP-hard problem \cite{AusielloBook1999}.

\begin{figure*}[b!t]
    \centering   \includegraphics[width=1.0\linewidth]{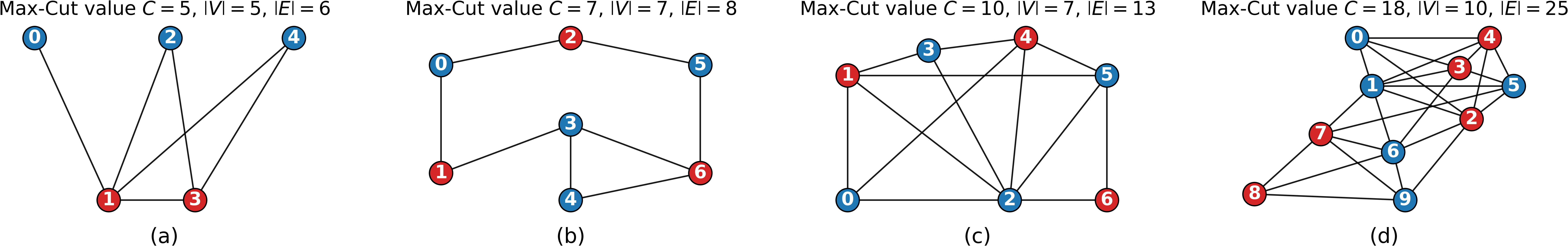}
    \caption{Examples of graphs together with one solution of Max-Cut problem for each. At the top of each example, we display the number $|V|$ of vertices, $|E|$ of edges, and $C$ of connecting edges for the optimal solution.}
    \label{GraphEx}
\end{figure*}

Despite the intense scrutiny of the Max-Cut problem, the importance of one simple fact was not emphasized strongly enough in the literature: The Max-Cut problem {\it always} has multiple solutions! This is simply because $V_1 \cup V_2= V_2 \cup V_1$ and, while classically these can be considered as identical solutions, quantum mechanically it implies that the spectrum of the Max-Cut Hamiltonian (see section~\ref{Sec:MaxCut}) is doubly degenerate. While this degeneracy can be attributed to the ``particle-hole" symmetry $|0\rangle \rightleftharpoons |1\rangle$, we found that there is a high probability for the Max-Cut problem to display four or more solutions (see section~\ref{Sec:TopoSF}). In such cases, the adiabatic spectral flow of the QAA looks as in Fig.~\ref{Schematic}, where we see spectral branches traversing the ground state spectral gap, an effect that is unavoidable due to a topological obstruction (see section~\ref{Sec:TopoSF}). The reality is that the {\it standard} formulation of the Adiabatic Theorem {\it never} applies to the Max-Cut problem.

Our main finding is that, despite the mentioned topological obstruction, the exactly implemented and converged adiabatic evolution performs flawlessly: In all instances we looked at, QAA correctly returns {\it all} existing solutions (see section~\ref{Sec:QAAPerformance})! This statement seems paradoxical because QAA always starts and ends with a single quantum state. However, the final state is always an entangled multi-qubit state with non-zero amplitudes for all Max-Cut solutions. Indeed, as we explain in section~\ref{Sec:Resolution}, the Adiabatic Theorem can be applied in two stages, first for the pair of projections $P_{\Gamma_1}$ and $P_{\Gamma_2}$, and second for the pair $P_{\Gamma_3}$ and $P_{\Gamma_4}$, with $\Gamma_i$'s as shown in Fig.~\ref{Schematic}. This assures us that $U_\tau(\tau,0)\psi(0)$ lands in the range of $P_{\Gamma_4}$ and, as such, in the manifold of all existing solutions. Furthermore, because there are no degeneracies in the bottom spectrum, except at the extreme end $s=0$, the dynamical part of the adiabatic evolution rapidly explores the range of $P_{\Gamma(s)}$ and creates the entanglement we mentioned. Probabilistically speaking, there is zero chance for the final state to display zero-amplitude for one or more solutions, and this is the mechanism by which QAA {\it simultaneously} detects all solutions of a Max-Cut problem.

Additionally, we investigated the performance of QAAs under realistic noise for four graphs displaying multiple solutions for the Max-Cut problem. The results confirm that the algorithms continue to identify correctly all solutions of the Max-Cut problem. This opens new directions for the applications of QAAs on existing quantum hardware.

The paper is organized as follows. In sections~\ref{Sec:MaxCut} and \ref{Sec:QAA}, we briefly review the Max-Cut problem in its QAA form, as well as its quantum circuit implementation in Qiskit environment. In section~\ref{Sec:TopoSF}, we develop a new strategy to compute the adiabatic spectral flow entirely inside the Qiskit environment, and use it to illustrate the flow for Max-Cut problems with multiple solutions. A simple topological invariant is introduced, whose non-trivial value denies the applicability of the standard Adiabatic Theorem. In section~\ref{Sec:QAAPerformance}, we analyze the performance of the QAAs, as simulated by Qiskit-Aer under various levels of noise. Section~\ref{Sec:Resolution} supplies an explanation of the newly found capacity of the QAAs to find the entire set of solutions in a single adiabatic cycle.

\begin{figure*}
    \centering   \includegraphics[width=1.0\linewidth]{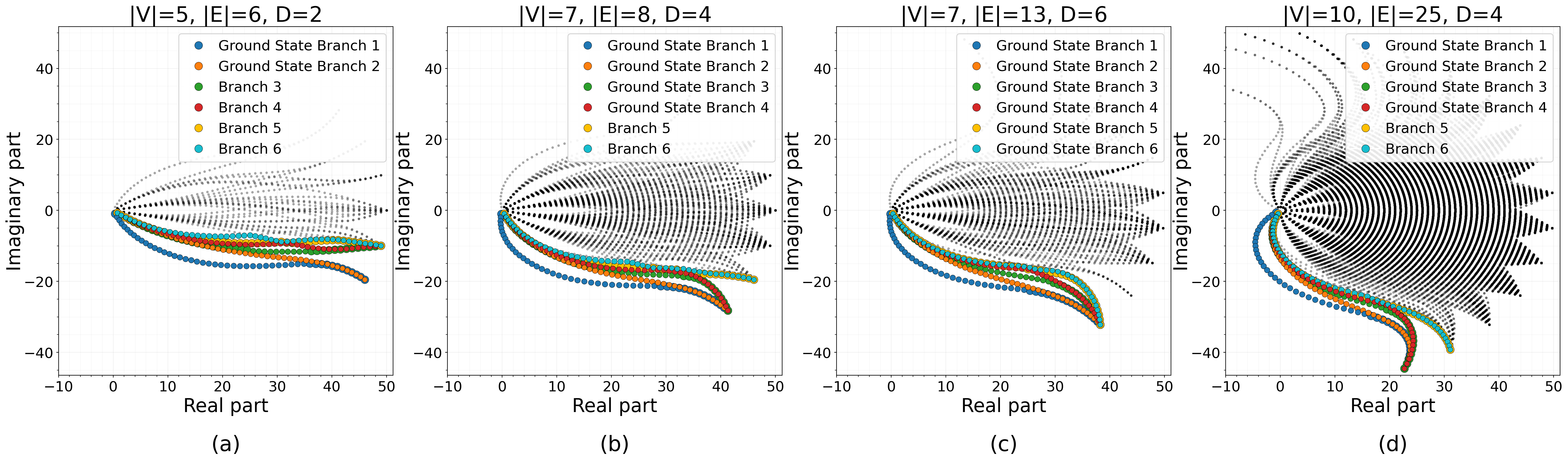}
    \caption{Spectral flow of $\Lambda(s) e^{\imath t H(s)}$, $s\in [0,1]$, $\Lambda(s)=20s$, generated by the Qiskit script from Listing~\ref{Script:2} on the graphs from Fig.~\ref{GraphEx}. The lowest six eigenvalues have been highlighted to facilitate their tracking, and the spectral branches landing in the ground state manifold are identified in the legends. At the top of each panel, we specify the degeneracy $D$ of the ground-levels.} 
    \label{TopoSF}
\end{figure*}

\section{The Max-Cut Problem} 
\label{Sec:MaxCut}

The Max-Cut problem involves graphs which we specify by their sets of vertices $V$ and edges $E$. The vertices will be enumerated in a rather arbitrary fashion $\{1,2, \ldots,|V|\}$, which is fixed from the beginning. The set $E$ of edges is a collection of pairs $(i,j)$ indicating which vertices are connected by an edge. Given such graph, the Max-Cut problem asks to find the optimal partition $V=V_1 \cup V_2$ of the vertices such that the number of edges connecting $V_1$ and $V_2$ is maximal. Examples of graphs and solutions of the Max-Cut problem are supplied in Fig.~\ref{GraphEx}.

Max-Cut problem is of interest in quantum computation because the partitions of a graph can be efficiently encoded in the quantum states of $|V|$ qubits placed at the vertices of the graph. Furthermore, the optimal partitions can be detected as the ground states of a specialized Hamiltonian. Indeed, any partition $V=V_1 \cup V_2$ can be encoded in a state $|v_1 \cdots v_{|V|}\rangle$, where the qubits inside $V_1$ are set in state 0 and the ones inside $V_2$ in state 1. Another key observation is that, if $(i, j) \in  E$, then $v_i + v_j - 2v_i v_j$ evaluates to 0 or 1, and 1 happens if and only if the edge $(i, j)$ connects $V_1$ and $V_2$. Thus, if $|v_1 \cdots v_{|V|}\rangle$ encodes a partition, the quantity
\[
C(\{v_i\}) = \sum_{(i,j)\in E} (v_i + v_j - 2 v_i v_j )
\]
computes the number of edges connecting the parts. Furthermore, if $X$, $Y$ and $Z$ denote Pauli's operators, then
\[
\tfrac{1}{2}(1-Z_i Z_j)|v_1 \cdots v_{|V|}\rangle = (v_i + v_j - 2v_iv_j)|v_1 \cdots v_{|V|}\rangle,
\]
and, as such, the quantum Hamiltonian
\[
H_{MC} = -\tfrac{1}{2}\sum_{(i,j)\in E}(1-Z_i Z_j)
\]
has the property that
\[
H_{MC}|v_1 \cdots v_{|V|}\rangle = - C(\{v_i\})\, |v_1 \cdots v_{|V|}\rangle.
\]
As a consequence, the solutions of the Max-Cut problem for a given graph are encoded in the ground state manifold of the quantum Hamiltonian $H_{MC}$. 

Since the cost function can be encoded in a quadratic Hamiltonian, Max-Cut problem belongs to the larger class of quadratic variational optimization problems, for which QAAs will formally be the same \cite{BlekosPR2024,GrangeAOR2024}. Thus, all our statements have a much broader applicability.

\section{Quantum Adiabatic Algorithm}
\label{Sec:QAA}

As explained in our introductory remarks, QAA is rooted in the Adiabatic Theorem. The standard formulation of the Adiabatic Theorem was thought to be sufficient for applications to Max-Cut problem, but that is certainly not the case due to the complications coming from the topological obstructions mentioned in our introductory remarks. However, regardless of the validity of the Adiabatic Theorem, the unitary adiabatic time evolution of a quantum system  is generally amenable  to quantum circuit implementations. Indeed, if $U_\tau(t,t')$ is  the time evolution operator for the time dependent Hamiltonian $H(t/\tau)$, that is, $U_\tau(t,t')$ is the solution of the operator equation
\begin{equation}
    \imath \partial_t U_\tau(t,t') = H(t/\tau) U_\tau(t,t'), \quad U_\tau(t',t') = I,
\end{equation}
then $\psi(\tau) = U_\tau(\tau,0) \psi(0)$ and $U_\tau (\tau,0)$ can be implemented by a sequence of unitary gates
\begin{equation}\label{Eq:UnitaryEvol1}
\begin{aligned}
    U_\tau(\tau,0) & = \prod_{n=1}^{N_t} U_\tau(n \Delta t, (n-1)\Delta t) \\
    & \approx \prod_{n=1}^{N_t} e^{\imath \Delta t \, H(\frac{n\Delta t}{\tau})},
    \end{aligned}
\end{equation}
where $N_t$ is a large number and $\Delta t = \tau/N_t$.


\begin{figure*}
    \centering
    \includegraphics[width=1.0\linewidth]{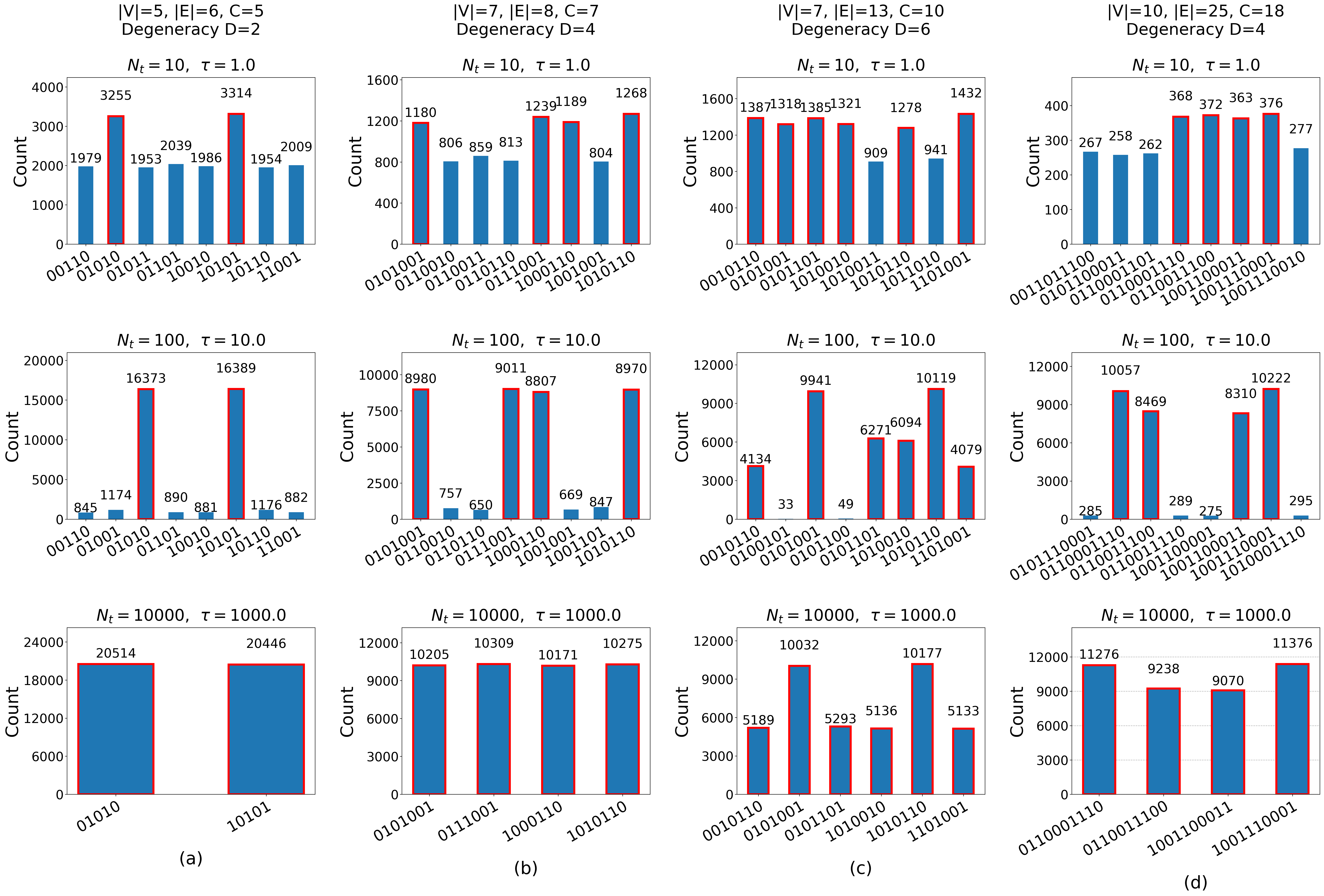}
    \caption{Outputs of the QAA algorithm for increasing number of time steps 
    ${\rm N_t} \in \{10, 100, 10000\}$ at fixed $\Delta t = 0.1$. Each (a-d) column corresponds to the matching (a-d) graph from Fig.~\ref{GraphEx}. The histograms were generated with $40,960$ shots and only the 8 most frequent bitstrings are displayed. The bitstrings corresponding to the solutions identified by a brute-force search are highlighted. Bitstrings are shown in MSB-first order as 
    $z = z_{n-1} z_{n-2} \cdots z_1 z_0$, where $z_0$ denotes qubit 0.}
    \label{Convergence_All}
\end{figure*}

For the Max-Cut problem on a graph, the strategy is to choose $H(s)$ to be a linear interpolation
\begin{equation}\label{Eq:Interpolation}
    H(s) = (1-s)H_M + s H_{MC},
\end{equation}
where $H_M$ is a simple Hamiltonian with known ground state, often called the mixer Hamiltonian 
\cite{BlekosPR2024}. A good choice is $H_M = - \sum_{i} X_i$, whose ground state is $\Hh^{\otimes V} |0,\ldots,0\rangle$, with $\Hh$ the Hadamard gate. With these choices, Eq.~\ref{Eq:UnitaryEvol1} can be further processed
\begin{equation}\label{Eq:UnitaryEvol2}
\begin{aligned}
    U_\tau(\tau,0) & \approx \prod_{n=1}^{N_t} e^{\imath \Delta t \, [(1-\frac{n}{N_t})H_M + \frac{n}{N_t} H_{MC}]} \\
    & \approx \prod_{n=1}^{N_t} e^{\imath \Delta t \, (1-\frac{n}{N_t})H_M} \, e^{\imath \Delta t \, \frac{n}{N_t} H_{MC}},
    \end{aligned} 
\end{equation}
and the terms seen above can be implemented by elementary gates \cite{Qiskit}:    
\begin{equation}\label{Eq:UnitaryEvol3}
    e^{\imath \beta H_M}= \prod_{i\in V} R_{X_i}(2\beta)
\end{equation}
and
\begin{equation}\label{Eq:UnitaryEvol4}
    e^{\imath \gamma H_{MC}}=\prod_{(i,j) \in E}R_{Z_i Z_j}(-2\gamma).
\end{equation}

A minimal Qiskit script generating the quantum circuit derived from Eqs.~\eqref{Eq:UnitaryEvol2}, \eqref{Eq:UnitaryEvol3} and \eqref{Eq:UnitaryEvol4} of the Max-Cut problem on graph (a) of Fig.~\ref{GraphEx}, is supplied below. To adapt the script for other graphs, one only needs to update the sets of vertices and edges of the graph. 
{\scriptsize
\begin{lstlisting}[language=Python,caption={ \ },label={Script:1}]
vertices=[0,1,2,3,4]
edges = [(0,1),(1,2),(1,3),(1,4),(2,3),(3,4)]
n_qubits = len(vertices)
N_t = 1000 # number of time steps 
dt = 0.1 # time interval
s_vals = np.linspace(0,1,N_t,endpoint=True)
qc = QuantumCircuit(n_qubits)
qc.h(range(n_qubits))  
for s in s_vals:
    b,g = (1-s)*dt,s*dt
    for q in vertices:
        qc.rx(2*b,q)
    for (i,j) in edges:
        qc.rzz(-2*g,i,j)
\end{lstlisting}
}

The performance of the quantum circuit will be analyzed in section~\ref{Sec:QAAPerformance}.
\section{Topological Spectral Flow}
\label{Sec:TopoSF}

Using the examples from Fig.~\ref{GraphEx}, we point out that:
\begin{itemize}
    \item The Max-Cut problem always has at least two solutions because $V_1 \cup V_2 = V_2 \cup V_1$.

    \item The graph in Fig.~\ref{GraphEx}(a) displays two solutions, the graphs in Figs.~\ref{GraphEx}(b,d) display four solutions, and the graph from Fig.~\ref{GraphEx}(c) displays six solutions for the Max-Cut problem!
    
    \item As one can see, multiple solutions can occur for a broad range of complexity, from the simple graph in Fig.~\ref{GraphEx}(b) to the fairly complex one in Fig.~\ref{GraphEx}(d).
\end{itemize}

\begin{figure*}
    \centering
    \includegraphics[width=1.0\linewidth]{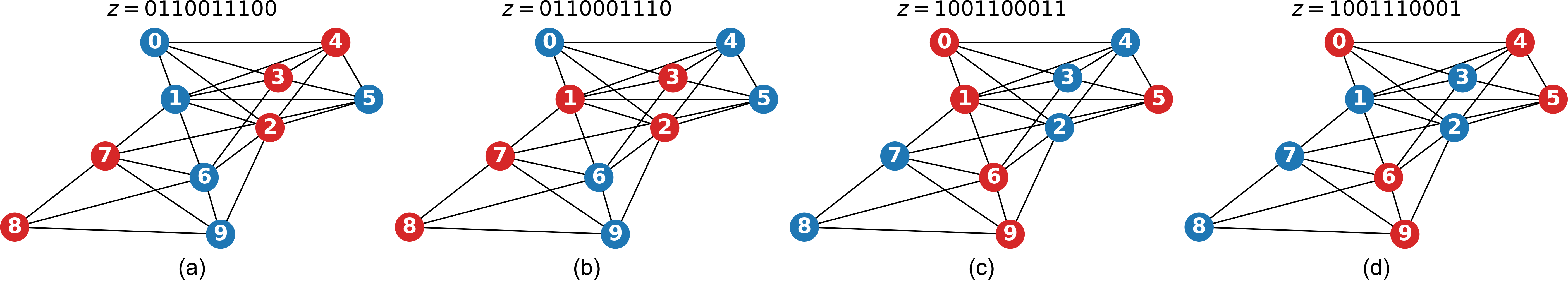}
    \caption{The complete set of solutions of the Max-Cut problem obtained via classical brute-force search for the graph from Fig.~\ref{GraphEx}(d). Each classical solution can be paired with a bitstring produced by the quantum algorithm (see Fig.~\ref{Convergence_All}(d)), demonstrating perfect agreement and validation of the QAA output.}
    \label{Fig:MC_Solutions}
\end{figure*}

Since we are dealing with finite graphs, the Hamiltonians appearing in the quantum formulation of the Max-Cut problem belong to the algebra of finite matrices $M_{N_V}(\CM)$, $N_V = 2^{|V|}$. Inside this algebra, a projection can be continuously deformed into another projection only if the two projections have equal rank. In other words, for this simple algebra, the rank is the complete topological invariant of a projection. Therefore, if $H_M$ and $H_{MC}$ display spectral projections of distinct ranks, the eigenvalues corresponding to these spectral projections cannot evolve into each other under an adiabatic deformation without touching other eigenvalues. In particular, this is certainly the case for the ground states of $H_M$ and $H_{MC}$, which are non-degenerate in the first case and at least doubly degenerate in the second case. The reality is that there is a topological obstruction which prevents the standard conditions of the Adiabatic Theorem to ever be satisfied for the Max-Cut problem. For this reason, the gap above the ground state of $H(s)$ necessarily closes for one or more values of $s$, and this will unavoidably be the case for all continuous deformations of $H_M$ into $H_{MC}$. For this reason, the evolution of the spectrum with the parameter $s$ is referred to as topological spectral flow.

In fact, we can associate a very simple topological invariant to the adiabatic flow. For this, we consider two points inside the spectral gaps of $H(0)$ and $H(1)$ and a continuous line connecting them (see the red dots and line in Fig.~\ref{Schematic}). We will define an intersection index between this continuous line and the spectral flow: If a spectral branch crosses the line from above, we count that intersection as $+1$ and, if it crosses from below, we count the intersection as $-1$ (see Fig.~\ref{Schematic}). The intersection index is the algebraic sum
\begin{equation}
    {\rm Index} = \# {\rm \ above \ crossings} - \# {\rm \ below \ crossings}
\end{equation}
which is independent of how we choose the line and on how we deform $H(0)$ into $H(1)$, as long as it is a continuous deformation. Indeed, upon deformations of the line and of the spectral flow, pairs of $\pm 1$ intersections can show up or disappear, but that cannot change the value of the index, which is always equal to ${\rm Rank}\, P_{\Gamma_4}-{\rm Rank}\, P_{\Gamma_1}$.

\begin{figure*}
    \centering
    \includegraphics[width=1.0\linewidth]{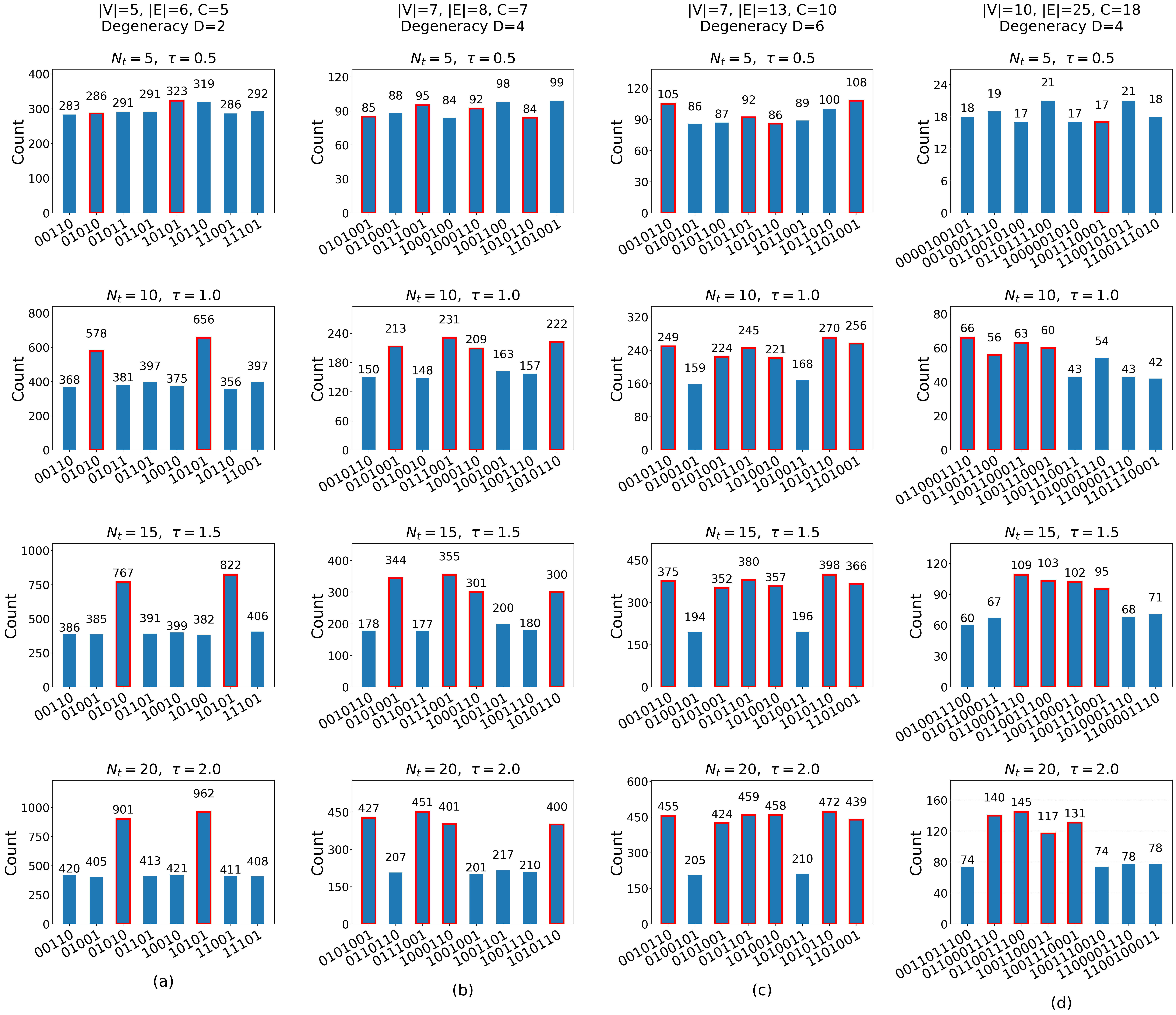}
    \caption{Same as Fig.~\ref{Convergence_All}, but simulated under an 
    optimistic IBM Heron r3 noise model corresponding to the 
    best-performing qubit couplings available on current 
    hardware~\cite{GambettaIBM2025,IBMQuantumHardware}. For parameters, see text. }
    \label{Fig:Convergence_Noisy_HeronR3_Opt}
\end{figure*}

Given these features, it will be desirable to map the adiabatic spectral flows of the Max-Cut problems, whenever that is feasible. Qiskit platform enables hybrid runs which blend classical evaluations with evaluations executed on quantum hardware. Furthermore, Listing~\ref{Script:1} demonstrates, among other things, Qiskit's efficiency when it comes to coding time evolutions of quantum spin Hamiltonians. As such, it is tempting to inquire if the adiabatic spectral flows can be mapped using a Qiskit script. We make the observation that, given any Hamiltonian $H$ with a spectrum rendered as usual on the real axis, its exponentiation as in the time evolution $e^{\imath t H}$ wraps its spectrum on the unit circle of the complex plane. If $t$ is small enough, the mapping is one-to-one and, as such, we can use the spectrum of the unitary operator $e^{\imath t H(s)}$ to map the flow of the spectrum with parameter $s$. Here, $t$ only plays the role of a scaling parameter: If $t$ is too large, the spectrum will wrap many times around the circle and, if $t$ is too small, then the spectrum is concentrated around 1 and is hard to visualize and examine. In our case, we fixed $t$ for the best visualization of the spectral flow.

We can generate the quantum circuit corresponding to $e^{\imath t H(s)}$ by using the Trotter expansion ($\Delta t = t/N_t$)
\begin{equation}\label{Eq:EH1}
\begin{aligned}
    e^{\imath t H(s)} & = \prod_{n=1}^{N_t} e^{\imath \Delta t \, H(s)} \\
    & \approx \prod_{n=1}^{N_t} e^{\imath \Delta t \, (1-s)H_M} \, e^{\imath \Delta t \, s H_{MC}}
    \end{aligned}
\end{equation}
and Eqs.~\eqref{Eq:UnitaryEvol3} and \eqref{Eq:UnitaryEvol4}. Qiskit can efficiently retrieve the unitary operator corresponding to a circuit and the spectrum of such unitary operator can be computed by classical means. A Qiskit script which computes and renders the spectra of $e^{\imath tH(s)}$ is reported below in the Listing~\ref{Script:2}. Note that, for visualization purposes, the spectra are scaled by a factor $\Lambda(s)$ to avoid overlapping. This is equivalent to computing the spectra of $\Lambda(s)e^{\imath t H(s)}$.

{\scriptsize
\begin{lstlisting}[language=Python,caption={ \ },label={Script:2}]
vertices=[0,1,2,3,4]
edges = [(0,1),(1,2),(1,3),(1,4),(2,3),(3,4)]
n_qubits = len(vertices)
N_s = 20 # sampling of s
N_t = 50  # number of time steps 
dt = 0.1
s_vals = np.linspace(0,1,N_s,endpoint=True) 
for s in s_vals:
    L_s = 20*s
    b,g = dt*(1-s),dt*s
    qc = QuantumCircuit(n_qubits)
    for kt in range(N_t):        
        qc.rx(2*b,vertices)
        for (i,j) in edges:
            qc.rzz(-2*g,i,j)
    U = Operator(qc).data
    eigvals,eigvecs = np.linalg.eig(U)
    x_real = np.real(L_s*eigvals)
    x_imag = np.imag(L_s*eigvals)
    plt.scatter(x_real,x_imag)
plt.show()
\end{lstlisting}
}

The above Qiskit script was used to generate Fig.~\ref{TopoSF}, which confirms the topological spectral flows predicted in the previous section for the graphs from Fig.~\ref{GraphEx}. Indeed, the lower branch seen in each panel represents the flow of the ground state eigenvalue and, in all cases, it is met by other branches at $s=1$ (see the highlights). Furthermore, the total number of branches meeting at $s=1$ is consistent with the number of solutions of the Max-Cut problem on the corresponding graphs. The complete Qiskit script used to generate Fig.~\ref{TopoSF} can be found at \cite{QiskitScripts}.

\begin{figure*}
    \centering
    \includegraphics[width=1.0\linewidth]{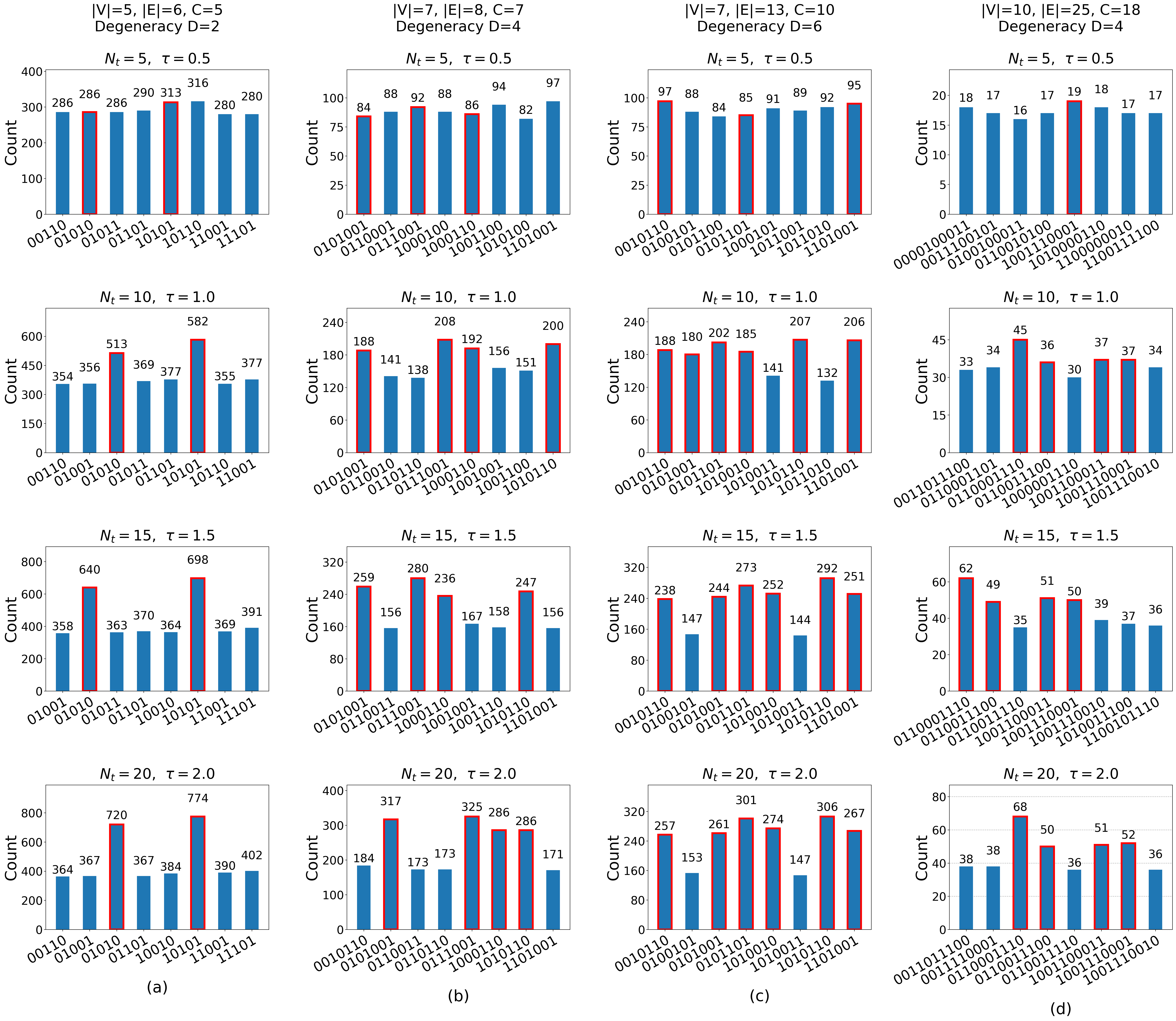}
    \caption{Same as Fig.~\ref{Convergence_All}, but simulated under a 
    realistic IBM Heron r2 noise model calibrated to median device 
    performance across the qubit register~\cite{IBMQuantumHardware,IBMQPUdocs}. For parameters, see text.} 
    \label{Fig:Convergence_Noisy_HeronR2_Avg}
\end{figure*}

\section{QAA Performance}
\label{Sec:QAAPerformance}

We tested the QAA algorithms as applied to the Max-Cut problem for the graphs from Fig.~\ref{GraphEx}, by running the script from Listing~\ref{Script:1} on the quantum simulator Qiskit-Aer with various levels of noise. The complete Qiskit scripts can be found at \cite{QiskitScripts}. 

The results of the simulations without noise are reported in Fig.~\ref{Convergence_All}, where (partial) histograms of multi-qubit measurements (aka shots), taken after the quantum circuit was applied on qubits initialized in the maximally entangled state, can be seen for three values of $\tau$. For clarity, the histograms display only the first 8 most frequent bitstrings. We recall that, to enter the adiabatic regime,  $\tau$ must be a number much greater than 1, while $\Delta t$ should be much smaller than 1. The latter was fixed at $0.1$ in these simulations. Since the depth of the QAA quantum circuit is proportional to $N_t=\tau/\Delta t$, we adjusted $\tau$ in the first row of Fig.~\ref{Convergence_All} such that the quantum circuit is amenable on existing quantum hardware. In the third row, we adjusted $\tau$ to achieve adiabatic convergence. The second row corresponds to an intermediate value and its sole purpose is to sample the convergence process. Additionally, we used a brute-force search algorithm to detect all solutions of the Max-Cut problems, and their corresponding bitstrings are highlighted in Fig.~\ref{Convergence_All}.

As one can see, already for $\tau = 1$ ($N_t=10$), all the solutions of the Max-Cut problems appear among the seen 8 most frequent bitstrings, for all our tested graphs. In the fully converged limit $\tau =1000$ ($N_t=10,000$), the histograms are entirely concentrated on the solutions of the Max-Cut problems, since the frequencies seen there add up exactly to the number of shots! This is an irrefutable evidence that the quantum adiabatic evolution of the ground state $\psi_0=\Hh^{\otimes V}|0\ldots 0\rangle$ of $H_M$ landed in the ground-state manifolds of $H_{MC}$, spanned by the solutions to the Max-Cut problems. For completion, these solutions are displayed in Fig.~\ref{Fig:MC_Solutions} for the graph from Fig.~\ref{GraphEx}(d). Additional tests confirm that this conclusion is robust against the chosen values of $\Delta t$ and $\tau$, as long as they correctly represent the adiabatic regime. 

In Fig.~\ref{Fig:Convergence_Noisy_HeronR3_Opt}, we tested QAA performance under an 
    optimistic IBM Heron r3 noise model corresponding to the 
    best-performing qubit couplings available on current 
    hardware~\cite{GambettaIBM2025,IBMQuantumHardware}. 
    Results are shown across four rows with increasing Trotter step count 
    $N_t \in \{5, 10, 15, 20\}$ at fixed time step $\Delta t = 0.1$, 
    giving total evolution times 
    $\tau = N_t \Delta t \in \{0.5, 1.0, 1.5, 2.0\}$, and 
    each column corresponds to one graph (a)--(d) from Fig.~\ref{GraphEx}. 
    Noise is modelled via Qiskit Aer all-qubit depolarizing channels 
    with single-qubit gate error 
    $p_\mathrm{1Q} = 1 \times 10^{-4}$~\cite{IBMQuantumHardware}, 
    two-qubit depolarizing error 
    $p_\mathrm{2Q} = 2 \times 10^{-3}$ per $R_{ZZ}$ gate 
    (reflecting the effective error of the native decomposition 
    $R_{ZZ} \rightarrow 2\,\mathrm{CZ}$ at the best-coupling 
    CZ error ${\approx}\,1 \times 10^{-3}$ on Heron 
    r3~\cite{GambettaIBM2025}), 
    and symmetric readout assignment error 
    $p_\mathrm{ro} = 8 \times 10^{-3}$~\cite{IBMQuantumHardware}. 
    Each circuit was executed with $8{,}192$ shots. 
    The correct Max-Cut solutions are clearly distinguishable above the 
    noise floor across all four graphs at all depth levels shown, 
    with peak prominence increasing monotonically with $N_t$. 
    Degradation with graph edge density is visible but mild, 
    reflecting the limited gate-error accumulation achievable 
    on the best available superconducting couplings. 

In Fig.~\ref{Fig:Convergence_Noisy_HeronR2_Avg}, we tested QAA performance under a 
    realistic IBM Heron r2 noise model calibrated to median device 
    performance across the qubit 
    register~\cite{IBMQuantumHardware,IBMQPUdocs}. 
    The results are organized and reported exactly as in Fig.~\ref{Fig:Convergence_Noisy_HeronR3_Opt}. 
    Noise, however, is modelled via Qiskit Aer all-qubit depolarizing channels 
    with single-qubit gate error 
    $p_\mathrm{1Q} = 2 \times 10^{-4}$~\cite{IBMQuantumHardware,IBMQPUdocs}, 
    two-qubit depolarizing error 
    $p_\mathrm{2Q} = 6 \times 10^{-3}$ per $R_{ZZ}$ gate 
    (reflecting the effective error of the native decomposition 
    $R_{ZZ} \rightarrow 2\,\mathrm{CZ}$ at the median Heron r2 
    CZ error ${\approx}\,3 \times 10^{-3}$~\cite{IBMQuantumHardware,IBMQPUdocs}), 
    and symmetric readout assignment error 
    $p_\mathrm{ro} = 1.5 \times 10^{-2}$~\cite{IBMQuantumHardware,IBMQPUdocs}. 
    Each circuit was executed with $8{,}192$ shots. 
    Despite hardware-realistic noise levels, the correct Max-Cut solutions 
    remain distinguishable above the noise floor across all four graphs 
    at all depth levels shown. Degradation in peak prominence with 
    increasing graph edge density reflects the growth in two-qubit gate 
    count per Trotter step, which constitutes the dominant 
    error-accumulation mechanism at current hardware fidelities. 
    
\section{Resolution and Conclusions}
\label{Sec:Resolution}

Despite the topological obstruction discovered by our work, the simulated QAA algorithms complete the Max-Cut tasks with somewhat unexpected accuracy: In all our tested cases, QAA was able to identify the {\it full} set of solutions! This calls for an explanation.

Let us recall that the topological obstruction stems from the distinct degeneracies of the ground-state manifolds of the mixer and Max-Cut Hamiltonians. This forces the gap above the ground state of the interpolating Hamiltonian $H(s)$ to eventually close and, as such, denying the application of the standard Adiabatic Theorem. However, by examining Fig.~\ref{TopoSF}, we see that the spectral flow is very particular because the spectral branches crossing the ground-state gap always do so from top to bottom. This particularity enables us to apply the standard Adiabatic Theorem in two steps: First, for the pair of spectral intervals $\Gamma_1 - \Gamma_2$ and, second, for the pair of spectral intervals $\Gamma_3-\Gamma_4$ from Fig.~\ref{Schematic}. If $\Gamma_2$ and $\Gamma_3$ are located at $s_0$ on the $s$-axis, then first application of the theorem assures us that $U_\tau(s_0\tau,0)\psi_0$ lands in the linear space $P_{\Gamma_2}(s_0) Q$, where $Q$ is the Hilbert space spanned by the qubit states. But $P_{\Gamma_2}(s_0) Q \subset P_{\Gamma_3}(s_0) Q$, and the second application of the theorem assures that, for any vector $\psi$ from $P_{\Gamma_3}(s_0) Q$, $U_\tau(\tau,s_0\tau)\psi$ lands in $P_{\Gamma_4}(s_0) Q$. Since the latter is exactly the ground-state manifold of $H_{MC}$, we can conclude that, indeed, the QAA circuit always maps the initial state $\psi_0$ into the ground-state manifold of $H_{MC}$, spanned by the qubit states $|{\bf v}^\alpha\rangle = |v^\alpha_1,\ldots,v^\alpha_{|V|}\rangle$, $\alpha =1,\ldots,D$, representing the Max-Cut solutions. Lastly, we recall that the adiabatic time-evolution rotates the vectors at every Trotter time step and, generally, the final outcome $U_\tau( \tau, 0)\psi_0$ is an unpredictable linear combination $\sum_{\alpha = 1}^{D} a_\alpha |{\bf v}^\alpha \rangle$. However, the probability for any of $a_\alpha$'s to be zero is null, and this explains why the histograms detect the entire set of solutions.

In conclusion, we have demonstrated and explained a mechanism by which QAAs are capable of detecting the entire set of solutions for a variational problem that lacks uniqueness, such as the Max-Cut problem. Although QAAs are among the most studied quantum algorithms, this feature was somehow overlooked. It definitely opens new directions for QAA applications. For example, for a fixed number of vertices, now one can efficiently investigate how often does the Max-Cut problem have a degeneracy $D$, $D=2,4,6,\ldots$.

Another fact revealed by our analysis is the promising performance of QAA under noise and for relatively small numbers of Trotter steps. This indicates that QAA applications of scales comparable to the ones presented here can be successfully run on existing hardware, which we plan to investigate in the near future.

\section*{Acknowledgment}
E.P. acknowledges support from the U.S. Army Research Office through contract W911NF-23-1-0127. P.S.J. acknowledges support from IBM through an IBM Masters Fellowship Award.

\end{document}